\documentclass[sigconf]{acmart}
\usepackage{xcolor}
\usepackage{xspace}
\usepackage{outlines}
\usepackage{graphicx}
\usepackage{float} 
\usepackage{multirow}
\usepackage{booktabs}
\usepackage{hyperref}
\usepackage{tcolorbox}
\usepackage{enumitem}
\usepackage[skip=0pt,size=small]{caption}

\AtBeginDocument{%
  }

\setcopyright{acmlicensed}
\copyrightyear{2025}
\acmYear{2025}
\acmDOI{XXXXXXX.XXXXXXX}

\acmConference[EASE 2025]{The 29th International Conference on Evaluation and Assessment in Software Engineering}{17–20 June, 2025}{Istanbul, Türkiye}

\acmISBN{978-1-4503-XXXX-X/18/06}

\begin{document}

    \title{Are We on the Same Page? Examining Developer Perception Alignment in Open Source Code Reviews}

    \author{Yoseph Berhanu Alebachew}
    \email{yoseph@vt.edu}
    \orcid{0000-0002-2922-4337}
    \affiliation{%
      \institution{Department of Computer Science, Virginia Tech}
      \city{Blacksburg}
      \state{Virginia}
      \country{USA}
    }

    \author{Minhyuk Ko}
    \email{minhyukko@vt.edu}
    \orcid{0009-0004-5473-4293}
    \affiliation{%
      \institution{Department of Computer Science, Virginia Tech}
      \city{Blacksburg}
      \state{Virginia}
      \country{USA}
    }
    
    \author{Chris Brown}
    \email{dcbrown@vt.edu}
    \orcid{0000-0002-6036-4733}
    \affiliation{%
      \institution{Department of Computer Science, Virginia Tech}
      \city{Blacksburg}
      \state{Virginia}
      \country{USA}
    }
    
    \renewcommand{\shortauthors}{Alebachew et al.}

\begin{abstract}
Code reviews are a critical aspect of open-source software (OSS) development, ensuring quality and fostering collaboration. This study examines perceptions, challenges, and biases in OSS code review processes, focusing on the perspectives of Contributors and Maintainers. Through surveys ($n=289$), interviews ($n=23$), and repository analysis ($n=81$), we identify key areas of alignment and disparity. While both groups share common objectives, differences emerge in priorities, e.g, with Maintainers emphasizing alignment with project goals while Contributors overestimated the value of novelty. Bias, particularly familiarity bias, disproportionately affects underrepresented groups, discouraging participation and limiting community growth. Misinterpretation of approach differences as bias further complicates reviews. Our findings underscore the need for improved documentation, better tools, and automated solutions to address delays and enhance inclusivity. This work provides actionable strategies to promote fairness and sustain the long-term innovation of OSS ecosystems.

\end{abstract}

\ccsdesc[500]{Software and its engineering~Programming teams}
\ccsdesc[500]{Software and its engineering~Open source model}
\keywords{Code Reviews, Bias in Software Development, Open Source Software (OSS), Inclusive Development, Contributors and Maintainers}
\received{31 January 2025}

\maketitle

\section{Introduction}

The code review process is a cornerstone of open-source software (OSS) development, ensuring that code changes meet quality standards, detecting defects early, and maintaining the integrity of the codebase~\cite{Benington1983, Patel2023}.
Unlike closed-source projects, where hierarchical teams oversee code quality, OSS relies on a decentralized, community-driven model~\cite{hippel2003open, feller2000framework}.  

In OSS, developers (referred to as \textit{\textbf{Contributors}} in this study) submit changes as pull requests, which are reviewed by a smaller group of developers with higher privileges, referred to here as \textbf{\textit{Maintainers}}. These Maintainers evaluate the submission and decide its outcome (i.e., acceptance or rejection).

The OSS model brings together developers with varying levels of expertise and experience, working collaboratively toward the shared goal of building a software. However, this diversity also introduces conflicts and biases in the code review process, which can undermine inclusivity and hinder the growth of OSS communities~\cite{Vasilescu2015}. Several studies show the existence of bias in the code review process~\cite{Huang2020,Terrell2016, pushback,McGrath2019, MurphyHill2023, sultana2023CodeReviews}.

Studies show that biases in code reviews, such as favoring known Contributors over new or underrepresented ones, can discourage continued participation and hinder collaboration~\cite{Tsay2014, Bird2015}. These biases not only result in inefficiencies, delays, and conflicts but also reduce the talent pool, limiting diversity and innovation in open-source projects~\cite{Nadia2016, Vasilescu2015}. Inclusive communities are more effective at solving complex problems and sustaining long-term growth. Addressing bias is therefore not just a fairness issue but a critical factor in fostering sustainable and innovative OSS ecosystems~\cite{Huang2018}.

However, not all pushbacks and conflicts in code review are caused by bias. Prior work suggests the frequency and impact of bias is often overestimated in human judgment~\cite{pronin}. Tsay et. al \cite{Tsay2014} also report that misunderstandings in OSS environments are often rooted in differences in expectations rather than actual biases. To effectively address the challenges posed by biased code reviews, it is crucial to differentiate between noise, which represents misunderstandings or misinterpretations, and signal, which reflects instances of true bias affecting the review process.

Building on the foundational work of Tversky and Kahneman, who described bias as a systematic deviation from objective standards, norms, or rationality in judgment or decision-making ~\cite{tversky1974judgment}, we aim to explore whether different stakeholders in OSS code reviews share a consensus on the objectives, key elements, and various aspects of the code review process. Understanding these differences is crucial, as misalignment in expectations or priorities can exacerbate misunderstandings, hinder collaboration, and potentially be understood as bias in the review process.

Towards this goal, this study seeks to answer the following research questions (RQs):
\begin{enumerate}[topsep=0pt,label=\textbf{RQ\arabic*}: ]
    \item To what extent do maintainers and contributors in open-source projects agree on the objectives, key success factors, challenges, and potential improvements in the code review process?
    \item How does alignment (or lack thereof) between maintainers and contributors influence perceptions of bias in open-source projects' code review?
    \item To what extent do developers believe the objectives, success factors, and challenges of the code review process reflected in project documentation, such as contribution guidelines?
\end{enumerate}

To answer these questions, we conducted online surveys ($n = 289$) and qualitative interviews ($n = 23$) with OSS developers experienced in contributing to and maintaining GitHub\footnote{\url{https://github.com/}} repositories to investigate OSS code review processes. Our study explores the dynamics between Contributors and Maintainers, addressing gaps in understanding bias, communication, and documentation.


To the best of our knowledge, this is the first study to examine the alignment of expectations, opinions, challenges, and suggestions in OSS code reviews across roles.
Our contributions include an analysis that identifies widespread misinterpretation of ``approach differences'' as bias, underscoring the need for clearer communication and documentation.

Additionally, our study reveals differing priorities between Contributors and Maintainers, offering actionable strategies to enhance alignment. While Maintainers prioritize aligning contributions with project goals, Contributors often emphasize innovation, underscoring the importance of bridging this gap through effective communication. By addressing both technical and interpersonal aspects of code reviews, we propose a roadmap for more inclusive and sustainable OSS ecosystems. Empirical evidence also shows bias disproportionately affects newer Contributors and underrepresented groups, aligning with prior research~\cite{Lee2020,McGrath2019}. Key contributions of this study include:

\begin{itemize}
    \item  Evidence that while groups align on core aspects of code review, subtle but significant differences exist, warranting further research to improve alignment.
    \item Empirical proof of misattributing approach differences as bias, calling for interventions to clarify perceptions and improve understanding.
\end{itemize}

\section{Related Work}

Prior research has extensively explored the dynamics of code reviews and collaboration in software development. Santos and Nunes demonstrated the effectiveness of peer code reviews in distributed development for improving code quality, collaboration, and maintainability, while emphasizing the need to balance technical rigor with constructive feedback~\cite{SantosNunes2022}. Similarly, Beller et al. highlighted the benefits of modern code reviews (MCRs) in enhancing code quality and fostering collaboration but noted their limitations in addressing architectural concerns, underscoring the need for complementary tools~\cite{Beller2021}.

The dual role of OSS code reviews in ensuring quality and fostering trust was further examined by Alami et al., who emphasized transparent communication and mentorship, while identifying challenges such as cultural differences and asynchronous communication~\cite{Alami2020}. Eisty and Carver focused on peer code reviews in research software, identifying critical factors such as clear objectives and domain-specific feedback, while advocating for better documentation and tailored review practices to address inconsistencies and time constraints~\cite{EistyCarver2022}.

Research has also investigated the broader roles of OSS maintainers and contributors. Dias et al. identified key attributes of maintainers, including decision-making and inclusivity, while highlighting challenges like burnout and contributor management. Their recommendations to leverage automation, mentorship, and clear guidelines underscore the importance of alignment between maintainers and contributors~\cite{Dias2021}. Constantino et al. emphasized the role of trust, effective communication, and goal alignment in successful OSS collaboration, while highlighting challenges such as onboarding and asynchronous workflows~\cite{Constantino2021}. Stray et al. similarly explored coordination in distributed teams, identifying tools, clear communication, and alignment as critical to overcoming cultural and temporal barriers~\cite{Stray2020}.

Other studies examined tools and guidelines in OSS processes. Wessel et al. found code review bots effective for automating tasks but noted their limitations in handling complex scenarios, advocating for customizable, context-aware tools~\cite{Wessel2021}. Elazhary et al. revealed gaps between contribution guidelines and actual practices in OSS projects, emphasizing the importance of actionable and up-to-date guidelines to reduce barriers and misaligned expectations~\cite{Elazhary2022}. While these studies examined different aspects of the open source code review process none explored the alignment of perceptions between Contributors and Maintainers.

\section{Methodology}
This study adopts a pragmatic epistemological stance, using mixed methods to address the research questions by balancing qualitative insights (from interviews and open-ended survey responses) with quantitative analysis (from surveys and repository data). We surveyed 289 participants from 81 open-source projects (102 Maintainers and 187 Contributors) and conducted follow-up interviews with 23 respondents (10 Maintainers and 13 Contributors) for deeper insights. The study received Institutional Review Board (IRB) approval, and the following sections outline the detailed process, from survey design to data analysis.

\subsection{Survey} 
\subsubsection*{\textbf{Design}}
The survey was designed to gather both quantitative and qualitative data on participants’ experiences with the code review process. Separate surveys were distributed to Maintainers and Contributors, with overlapping questions to allow for a comprehensive comparison. The surveys included multiple-choice, Likert-scale, and open-ended questions to capture nuanced insights.

The surveys had three sections: demographic information, background details (e.g., OSS experience, responsibilities, and code review/contribution experience), and role-specific questions on perceptions of the code review process, such as objectives, resources, and biases. Free-response options were included to capture any responses beyond predefined categories. Some questions were conditional; for example, participants who answered “yes'' to experiencing bias in code review were prompted to describe examples and resolutions.

\subsubsection*{\textbf{Validation}}
We validated the survey and interview questions through a pilot study and expert review with nine software engineering researchers. The pilot study involved six participants (three Maintainers and three Contributors) to identify ambiguities or clarity issues. Based on feedback from the pilot and expert review, we revised the questions, rewording complex terms and adding optional responses for questions that might not apply universally (e.g., project-specific review guidelines).

Following these revisions, the final surveys were distributed via the online platform QuestionPro\footnote{\url{https://questionpro.com}} following the procedure in section \ref{sec:repositories}. Participants were contacted by email and offered a \$10 Amazon gift card as an incentive. The complete set of survey questions is included in the supplementary material~\cite{supplemental}.

\subsection{Interview}
To gain a deeper understanding of the survey responses, follow-up semi-structured interviews were conducted with survey participants who expressed willingness to take part. Participants were contacted via their preferred contact method to schedule a 30\textasciitilde{}45-minute Zoom\footnote{\url{https://zoom.us}} interview. The interviews followed a predefined guide, included in the supplementary material~\cite{supplemental}, which was validated through a pilot study and expert reviews. Interviews were conducted by the first author and recordings were transcribed using Zoom’s built-in feature. To ensure accuracy, we performed another set of transcriptions using Whisper.cpp\footnote{\url{https://github.com/ggerganov/whisper.cpp}},and the discrepancies between the two versions were manually reconciled.
\subsection{Participant Recruitment}\label{sec:repositories}


To identify participants, we used the GitHub API\footnote{\url{https://api.github.com}} to filter the top (number of stars) 1,000 repositories\footnote{As of Feb 19, 2024, 10:55 AM ET} based on specific criteria. These criteria included repositories primarily comprising source code, being owned by organizations (defined as GitHub accounts classified as organizations), having at least three Maintainers, and using English as the primary language. The focus on English-language projects introduces potential bias, as it may exclude smaller or non-English-speaking projects that could exhibit different dynamics in their code review processes. We specifically targeted repositories with active contributions and discussions, prioritizing projects with at least three Maintainers to ensure a minimum level of community involvement.


Following this filtering, we identified and contacted 5,262 Maintainers and Contributors of these projects. Recruitment materials were shared with GitHub users whose email addresses were publicly available. We received survey responses from 289 participants (102 Maintainers and 187 Contributors) representing 81 repositories, resulting in a response rate of 5.5\%. The response rate is comparable to other surveys in software engineering \cite{smith2013improving} due to common challenges reported in previous studies~\cite{Khalid2024}
.


These 81 repositories, with an average age of just over 9 years , span 538 topics  and 20 programming languages. Table \ref{tab:repo_statistics} summarizes key statistics for these repositories, including forks, stars, watchers, and repository sizes. The complete list of repositories is provided in our supplemental materials~\cite{supplemental}.



\begin{table}[htbp!]
    \centering
    \footnotesize
    \caption{Repository Popularity Summary Statistics}
    \resizebox{0.45\textwidth}{!}{
        \begin{tabular}{lrrrrr}
            \toprule
             & \textbf{No. Forks}  & \textbf{Size (MB)} & \textbf{Stars} & \textbf{Watchers} \\
            \midrule
            \textbf{Average} & 8402.41 & 472.58 & 51357.32 & 49918.63 \\
            \textbf{Max} & 39062 &  8177.44 & 207525 & 206888 \\
            \textbf{Median} & 4339 & 148.32 & 35507 & 34291 \\
            \textbf{Min} & 904 & 0.58 & 24758 & 24479 \\
            \textbf{std($\sigma$)} & 9007.72 & 1076.39 & 34362.11 & 34013.75 \\
            \textbf{Total} & 680595 & 38278.69 & 4159943 & 4043409 \\
            \bottomrule
        \end{tabular}
    }
    \label{tab:repo_statistics}
\end{table}



Initially, we contacted Maintainers of the identified repositories to recruit survey participants. Subsequently, we targeted the Contributors of these repositories. This strategy aimed to maximize the validity of our comparison by ensuring that both groups of participants were involved in similar projects. As shown in Table \ref{tab:participant_stat}, similar demographic groups dominate among both Contributor and Maintainer participants.

\begin{table}[htbp!]
    \centering
    \caption{Summary of Participants' Demographics}
    \resizebox{0.45\textwidth}{!}{
                \begin{tabular}{l p{3cm} |r| r| r| r|}
            \cmidrule(r){3-6}
            \multirow{2}{*}{} & & \multicolumn{2}{c|}{Contributors} & \multicolumn{2}{c|}{Maintainers} \\
            \cmidrule(r){3-4} \cmidrule(r){5-6}
            & & Count & Percentage & Count & Percentage \\
            \midrule
            \multirow{4}{*}{\rotatebox{90}{\parbox{2cm}{\centering Age Group}}}
                & 18-24 & 35 & 18.72\% & 23 & 22.55\% \\
                & 25-34 & 89 & 47.59\% & 53 & 51.96\% \\
                & 35-44 & 44 & 23.53\% & 17 & 16.67\% \\
                & 45-54 & 11 & 5.88\% & 9 & 8.82\% \\
                & 55-64 & 7 & 3.74\% & - & - \\
                & Above 64 & 1 & 0.53\% & - & - \\
            \midrule
            \multirow{4}{*}{\rotatebox{90}{\centering Gender}}
                & Male & 176 & 94.12\% & 93 & 91.17\% \\
                & Female & 4 & 2.14\% & 7 & 6.86\% \\
                & Other & 4 & 2.14\% & - & - \\
                & Prefer not to say & 3 & 1.60\% & 2 & 1.96\% \\
            \midrule
            \multirow{5}{*}{\rotatebox{90}{\parbox{2cm}{\centering Professional Experience}}}
                & < 1  &  7 &  3.74\% & 8 & 7.84\% \\
                & 1-3  & 35 & 18.72\% & 17 & 16.67\% \\
                & 4-6  & 37 & 19.79\% & 25 & 24.51\% \\
                & 7-10 & 33 & 17.65\% & 20 & 19.61\% \\
                & >10  & 75 & 40.11\% & 32 & 31.37\% \\
            \midrule
            \multirow{5}{*}{\rotatebox{90}{\parbox{2cm}{\centering OSS Experience (Years)}}}
                & < 1  & 17 &  9.09\% & 4 & 3.92\% \\
                & 1-3  & 43 & 22.99\% & 26 & 25.49\% \\
                & 4-6  & 48 & 25.67\% & 37 & 36.27\% \\
                & 7-10 & 40 & 21.39\% & 14 & 13.73\% \\
                & >10  & 39 & 20.86\% & 21 & 20.59\% \\
            \midrule
            
            \multirow{5}{*}{\rotatebox{90}{\parbox{2cm}{\centering No Projects}}}
                & 1-3  & 45 & 24.06\% & 37 & 36.27\% \\
                & 3-5  & 52 & 27.81\% & 31 & 30.39\% \\
                & 6-10  & 37 & 19.79\% & 15 & 14.71\% \\
                & 11-20 & 15 & 8.02\% & 3 & 2.94\% \\
                & >20  & 38 & 20.32\% & 16 & 15.69\% \\
            \midrule
            \multirow{8}{*}{\rotatebox{90}{\parbox{2cm}\centering Race/Ethnicity}} 
                & Asian & 57 & 30.48\% & 29 & 28.43\% \\
                & Black or African American & 2 & 1.07\% & - & - \\
                & Multiracial & 15 & 8.02\% & 9 & 8.82\% \\
                & Native American or Alaska Native & 13 & 6.95\% & 2 & 1.96\% \\
                & Native Hawaiian or Other Pacific Islander & 1 & 0.53\% & - & - \\
                & Other & 9 & 4.81\% & 3 & 2.94\% \\
                & Prefer not to say & 86 & 45.99\% & 56 & 54.90\% \\
                & White & 4 & 2.14\% & 3 & 2.94\% \\
           \midrule
           \multirow{6}{*}{\rotatebox{90}{\parbox{2cm}{\centering Education Level}}} 
                & High school or equivalent & 30 & 16.04\% & 15 & 14.71\% \\
                & Some college/no degree & 15 & 8.02\% & 9 & 8.82\% \\
                & Associate degree & 7 & 3.74\% & 1 & 0.98\% \\
                & Bachelor’s degree & 77 & 41.18\% & 55 & 53.92\% \\
                & Master’s degree  & 46 & 24.60\% & 17 & 16.67\% \\
                & Doctorate or higher & 12 & 6.42\% & 5 & 4.90\% \\
            \midrule
            \multirow{7}{*}{\rotatebox{90}{\parbox{3cm}{\centering Participation Frequency}}} 
                & More than once a day & 14 & 7.47\% & 16 & 15.69\% \\
                & Daily  & 12 & 6.42\% & 11 & 10.78\% \\
                & Several times a week  & 32 & 17.11\% & 25 & 24.51\% \\
                & Weekly & 22 & 11.76\% & 18 & 17.65\% \\
                & Bi-Weekly  & 9 & 4.81\% & 6 & 5.88\% \\
                & Monthly  & 39 & 20.86\% & 11 & 10.78\% \\
                & Rarely  & 59 & 31.55\% & 14 & 13.72\% \\
                & Depends & - & - & 1 & 0.98\% \\
            \bottomrule
        \end{tabular}  
    }
    \label{tab:participant_stat}
\end{table}

\subsection{Participant Background}
Participants provided demographic and professional background information, including age, gender, professional and OSS experience, and project role. As indicated in Table \ref{tab:participant_stat}, respondents were predominantly male, aligning with previous findings \cite{Trinkenreich2022} that open-source software communities are male-dominated. Nearly half of the participants in both groups were aged 25–34. Similar parallels were observed in other demographics, such as ethnic/racial background and highest educational level achieved.


Out of 102 Maintainers surveyed, 92 individuals (approximately 90\%) described their role as ``Developer/Engineer''. Additionally, 36 Maintainers (about 35\%) reported having assumed a team leader or management role. What's more, seven Maintainers (approximately 7\%) are involved in Quality Assurance, another seven (7\%) work in DevOps or SysAdmin roles, five (about 5\%) identify as UX/UI Designers, and five (5\%) hold the position of CTO or CIO. Notably, 31 Maintainers (around 30\%) indicated that they fulfill two roles, while eight have three or more responsibilities.

The selection of participants skewed toward experienced Contributors and Maintainers, with most participants having more than three years of experience in OSS. The majority of Maintainers reported shared decision-making responsibilities in code reviews, with some acting as the primary decision-makers, particularly in high-traffic repositories. The potential impact of professional experience on perceptions of bias was explored in later sections of the analysis. Among Contributors, 151 individuals (80.75\%) indicated their primary contribution as ``Code Contribution''. This was followed by ``Bug Reporting and Testing'' and ``Translation and Localization'', selected by 19 (10.16\%) and 6 (3.21\%) respondents, respectively. When asked, ``Which sort of contributions do you make second most frequently?'', responses were more diverse. The most common response was ``Bug Reporting and Testing'', selected by 61 (32.62\%) participants, followed by ``Documentation'' (27.27\%). A smaller percentage, 14.97\%, indicated their second most frequent contribution as ``Community Support''.

Maintainers were also asked about their level of responsibility as reviewers in code reviews. The findings revealed that ``Shared decision making'' was the most typical level of responsibility, with 55 (53.92\%) respondents selecting this option. This was followed by ``Primary decision maker'' (21.57\%, 22 respondents). A slightly smaller portion, 17 (16.67\%), indicated that they typically hold an advisory role in code reviews, while eight (7.84\%) reported participating as observers.

Additionally, respondents ranked their perceived primary responsibilities as code Maintainers in order of importance (i.e., ranking out of six options). The top-ranked responsibility was ``Ensuring code quality and standards'', identified as the highest priority by 43.14\% of participants. This was followed by ``Overseeing Project Direction'', ranked as the primary responsibility by 23.53\% of respondents. The third most important responsibility, noted by 13.73\% of respondents, was ``Managing code-base and documentation''. The variation in rankings, presented in Table \ref{tab:responsibility_ranking}, highlights the diverse roles and priorities among code Maintainers.

\begin{table}[htbp!]
    \centering
    \caption{\centering \footnotesize{Self Ranking of Important Responsibility as Maintainer}}
    \resizebox{0.45\textwidth}{!}{
    \begin{tabular}{lrrrrrr}
        \toprule
         & \multicolumn{6}{c}{\textbf{Rank}} \\
        \cmidrule(r){2-7}
        \textbf{Responsibility} & \textbf{1} & \textbf{2} & \textbf{3} & \textbf{4} & \textbf{5} & \textbf{6} \\
        \midrule
        Finalizing Code Decisions & 5.88 & 16.67 & 25.49 & 17.65 & 15.69 & - \\
        Ensuring Code Quality and Standards & 43.14 & 18.63 & 14.71 & 12.75 & 3.92 & - \\
        Managing Code-Base and Documentation & 13.73 & 20.59 & 15.69 & 17.65 & 14.71 & - \\
        Overseeing Project Direction & 23.53 & 14.71 & 15.69 & 14.71 & 10.78 & 0.98 \\
        Mentoring or Guiding Contributors & 6.86 & 21.57 & 18.63 & 18.63 & 23.53 & - \\
        Other & 0.98 & - & - & 0.98 & 0.98 & 50.00 \\
        \bottomrule
    \end{tabular}}
    \small $1 = Highest Rank, 6 = Lowest Rank$
    \label{tab:responsibility_ranking}
\end{table}

\subsection{Data Analysis}
All survey responses were anonymized prior to analysis. Qualitative responses were open-coded by the first and second authors using the grounded theory approach~\cite{glaser1967grounded}, identifying themes such as perceived bias, review objectives, and communication issues. Coding disagreements were resolved through discussion until consensus was reached, removing the need for formal inter-coder reliability measures. Final codes and descriptions are available in the supplementary materials~\cite{supplemental}.

We applied the Spearman correlation coefficient~\cite{spearman1904proof} to assess associations between group responses. This non-parametric test, robust to outliers and suitable for skewed data~\cite{gibbons1993nonparametric}, assumes ordinal, interval, or ratio data and monotonic relationships. These assumptions were met, as the survey data used Likert scales and exploratory analysis confirmed monotonic trends. Spearman’s correlation quantified alignment between Contributors and Maintainers.

The Chi-Square Test of Independence~\cite{pearson1900x2} was used to evaluate group differences. Assumptions, including categorical data, independent observations, and expected frequencies $\geq 5$ per contingency table cell, were addressed by grouping low-frequency responses into an ``Other'' category and excluding cases with frequencies $<5$~\cite{agresti2018statistical}. Post-hoc z-scores from residuals identified statistically significant categories ($p<0.05$)~\cite{haberman1973residuals}.

\section{Results}

\subsection{RQ1: Objectives, Key factors, Challenges and Improvements in Code Review}

To evaluate the alignment between Contributors and Maintainers on perceived objectives of the code review process, factors influencing pull request acceptance, and reported challenges and areas for improvement in open-source projects, we conducted a survey with open-ended questions. These questions enabled participants to express their views, facilitating a comparison of perspectives between the two groups. After open coding the responses, we calculated the percentage of responses corresponding to each identified theme within each group. A Spearman correlation analysis, detailed in Table \ref{tab:correlation-and-chi-square}, revealed a general correlation between the responses.

\begin{table}[htbp!]
    \centering
    \caption{\small Correlation and Chi-Square results for objectives, key factors, challenges, and improvements in OSS code review.}
    \resizebox{0.45\textwidth}{!}{
        \begin{tabular}{|p{1.65cm}|p{.5cm}|p{.75cm}|p{4cm}|}
            \hline
            \small \textbf{Category} & \small \textbf{Corr.}  & \small \textbf{$\chi^2$} & \small \textbf{Responses with Deviation} \\
            \hline
            \textbf{Objectives} & 0.96 & 195.02 & Project Goal, Maintainability, Security, Documentation, Community Building, Conciseness\\
            \hline
            \textbf{Key Factors} & 0.94 & 111.65 & Novelty, Rationale \\
            \hline
            \textbf{Challenges} & 0.71 & 362.13 & Expectation Misalignment, Volume, Lack of Documentation, Unclear Expectations, Tooling, Inconsistent Reviewer Experience, Bias \\
            \hline
            \textbf{Improve-ments} & 0.72 & 118.14 & Better Tools and Automation, Enhanced Collaboration, Enhanced Communication \\
            \hline
        \end{tabular}
    }
    \label{tab:correlation-and-chi-square}
    \small \textbf{Note}: All results are statistically significant ($p$-value $< 0.05$)

\end{table}

To uncover subtle yet significant differences in responses, we employed a Chi-Square test, which confirmed the presence of such disparities. A subsequent post-hoc test with adjusted residuals identified the specific areas contributing most to these misalignments, as shown in Table \ref{tab:correlation-and-chi-square}. This indicates that while there is general agreement on many aspects, nuanced discrepancies in opinions persist. All Chi-Square test results in Table \ref{tab:correlation-and-chi-square} have $p$-value $< 0.05$.

The discrepancies revealed by the Chi-Square test, further explored through post-hoc analysis with adjusted residuals, highlight fundamental differences in prioritization and perspective between Contributors and Maintainers, as detailed in Tables \ref{tab:objectives}, \ref{tab:key_factors}, \ref{tab:challenges}, and \ref{tab:improvements}.

\subsubsection*{\textbf{Objectives:}} 
Our open coding identified 13 codes for objectives of the code review process, along with an additional category for responses that did not fit into any other. As noted earlier, responses were generally aligned. Both Contributors and Maintainers emphasized ensuring ``Correctness'' (52.94\%) and ``Quality'' (42.21\%) of contributions as primary objectives. These align with prior research on developers' motivations for code reviews~\cite{Bacchelli2013}, highlighting a shared commitment to high standards and reliable code. Contributors also emphasized adherence to project standards and knowledge sharing, while Maintainers focused more on aligning contributions with overall project goals. Table \ref{tab:objectives} presents the top ten reported objectives for each respondent group.

Our post-hoc analysis reveals that while both groups prioritize similar aspects, the emphasis placed on certain objectives varies significantly. For example, a larger proportion of Maintainers consider aligning contributions with overall ``Project goals'' (31.37\%) as a key objective compared to Contributors (13.37\%). Other statistically significant differences include objectives such as ensuring ``Maintainability'' (11.76\%), ``Security'' (11.07\%), and proper ``Documentation'' (6.57\%). Objectives like ``Community Building'' (5.88\%) and ``Conciseness'' (2.77\%) were mentioned by fewer participants overall.

These differences were echoed in follow-up interviews. Several Maintainers emphasized aligning contributions with project goals. Participant M83 noted, \textit{``a contribution that often we will turn away ... is not aligned with what we want with this project.''} The significance of security was also stressed, with Participant M61 remarking, ``\textit{Security is often overlooked by contributors, but it’s a non-negotiable priority for any review.}'' Regarding documentation, Participant M65 commented, ``\textit{Proper documentation ensures that others can build upon the work without needing extensive clarifications.}''

\begin{table}[htbp!]
    \centering
    \caption{\small{Top 10 Reported Objectives of Code Review }}
        \begin{tabular}{|l|c|c|c|}
            \hline
            \textbf{Category} & \textbf{C (\%)} & \textbf{M (\%)} & \textbf{All (\%)} \\ \hline
            Correctness        & 57.75                     & 44.12                     & 52.94             \\ \hline
            Quality            & 40.64                     & 45.10                     & 42.21             \\ \hline
            Standard           & 24.60                     & 27.45                     & 25.61             \\ \hline
            Knowledge Sharing  & 20.86                     & 27.45                     & 23.18             \\ \hline
            Project Goal       & 13.37                     & 31.37                     & 19.72             \\ \hline
            Maintainability    & 11.23                     & 12.75                     & 11.76             \\ \hline
            Security           & 11.76                     & 9.80                      & 11.07             \\ \hline
            Documentation      & 7.49                      & 4.90                      & 6.57              \\ \hline
            Community Building & 5.35                      & 6.86                      & 5.88              \\ \hline
            Conciseness        & 3.74                      & 0.98                      & 2.77              \\ \hline
        \end{tabular}
    \label{tab:objectives}
     \small $C = Contributors;  M=Maintainers$
\end{table}

\begin{tcolorbox}[colback=gray!10, colframe=gray!80, title=Key Finding]
Perception of the objectives are mostly aligned between Maintainers and Contributors, but the former place a stronger emphasis on aligning contributions with the overarching project goals compared to the latter. 
\end{tcolorbox}

\subsubsection*{\textbf{Key Factors:}} 
Similarly, both groups identified 16 key factors with comparable themes, though their frequency differed. Our analysis highlights the critical factors participants identified for ensuring a pull request is accepted. As shown in Table \ref{tab:key_factors}, both Contributors and Maintainers prioritized ``Correctness'' (38.41\%), followed by ``Standards'' (34.26\%), and proper ``Documentation'' (31.49\%) as the most important factors.  These findings are supported by follow-up interviews, where participants emphasized these aspects. Participant M4 stated, ``Ensuring correctness is non-negotiable—it’s the baseline for any acceptable pull request.''  

While both groups consider similar factors as determinants for pull request acceptance, Contributors tend to overestimate the importance of ``Novelty'' (11.23\% vs 1.96\%) and underestimate the significance of ``Rationale'' (3.74\% vs 12.75\%). These differences are statistically significant ($p$-value $< 0.05$), suggesting that Contributors may not fully appreciate the importance of clearly communicating the rationale behind their contributions and design decisions. 

Insights from follow-up interviews further highlight this gap. Participant M61 observed, ``\textit{Contributors often focus on how new or innovative their ideas are but fail to explain why their changes are necessary or how they align with the project goals.}'' Similarly, Participant M63 emphasized, ``\textit{Without a clear explanation of the reasoning behind a change, it’s difficult for maintainers to assess its relevance and impact.}'' Some Contributors acknowledged the challenge, with Participant M74 noting, ``\textit{It's not always obvious what maintainers expect in terms of rationale, and this creates unnecessary friction.}''

\begin{table}[htbp!]
    \caption{\small{Top 10 Reported Key Factors for Pull Request Acceptance}}
        \begin{tabular}{|l|c|c|c|}
            \hline
            \textbf{Category} & \textbf{C (\%)} & \textbf{M (\%)} & \textbf{All (\%)} \\ \hline
            Correctness        & 40.11                     & 35.29                     & 38.41             \\ \hline
            Standard           & 32.62                     & 37.25                     & 34.26             \\ \hline
            Documentation      & 33.69                     & 27.45                     & 31.49             \\ \hline
            Quality            & 29.95                     & 34.31                     & 31.49             \\ \hline
            Communication      & 27.27                     & 24.51                     & 26.30             \\ \hline
            Project Goal       & 19.79                     & 28.43                     & 22.84             \\ \hline
            Conciseness        & 16.58                     & 14.71                     & 15.92             \\ \hline
            Novelty            & 11.23                     & 1.96                      & 7.96              \\ \hline
            Rationale          & 3.74                      & 12.75                     & 6.92              \\ \hline
            Responsiveness     & 5.35                      & 2.94                      & 4.50              \\ \hline
        \end{tabular}
    \label{tab:key_factors}
    \small $C = Contributors;  M=Maintainers$
\end{table}

\begin{tcolorbox}[colback=gray!10, colframe=gray!80, title=Key Finding]
Both groups prioritized ``Correctness,'' ``Standards,'' and ``Documentation'' for pull request acceptance, but Contributors overvalued ``Novelty'' and undervalued ``Rationale,'' highlighting differences in emphasis ($p$-value $< 0.05$).
\end{tcolorbox}

\subsubsection*{\textbf{Challenges:}}
We asked participants about the major challenges in the current code review process and their suggestions to determine whether Contributors and Maintainers share a common understanding. As shown in Table \ref{tab:challenges}, several key patterns emerged in the challenges perceived by both groups in the OSS code review process. Both groups identified ``Reviewer Responsiveness'' (29.07\%) as a major issue, with many respondents highlighting delays or inconsistencies in review responses. ``Communication Issues'' (9.34\%) also emerged as a common concern, emphasizing the critical but often insufficient role of effective communication.

\begin{table}[htbp!]
    \centering
    \caption{Reported Challenges in the Code Review Process}
    \resizebox{0.45\textwidth}{!}{
        \begin{tabular}{|l|c|c|c|}
            \hline
            \textbf{Category} & \textbf{C (\%)} & \textbf{M (\%)} & \textbf{All (\%)} \\ \hline
            Reviewer Responsiveness       & 29.41                     & 28.43                     & 29.07             \\ \hline
            Communication Issues          & 8.56                      & 10.78                     & 9.34              \\ \hline
            Expectation Misalignment      & 6.42                      & 9.80                      & 7.61              \\ \hline
            Volume                        & 5.88                      & 9.80                      & 7.27              \\ \hline
            Lack of Documentation         & 5.88                      & 2.94                      & 4.84              \\ \hline
            Code Quality Concerns         & 4.81                      & 3.92                      & 4.50              \\ \hline
            Unclear Expectations          & 4.81                      & 3.92                      & 4.50              \\ \hline
            Miscellaneous                 & 2.14                      & 4.90                      & 3.11              \\ \hline
            Tooling                       & 1.07                      & 5.88                      & 2.77              \\ \hline
            Inconsistent Reviewer Experience & 2.67                   & 0.98                      & 2.08              \\ \hline
            Bias                          & 1.07                      & 2.94                      & 1.73              \\ \hline
            Contributor Responsiveness    & 2.14                      & 0.98                      & 1.73              \\ \hline
        \end{tabular}
    }
    \label{tab:challenges}
    \small $C = Contributors;  M=Maintainers$
\end{table}

However, notable differences were observed between the two groups. Contributors were more concerned about ``Lack of Documentation'' (5.88\% vs 2.94\%) and ``Inconsistent Reviewer Experience'' (2.67\% vs 0.98\%), suggesting challenges in understanding review expectations and variability in review quality. Maintainers, on the other hand, reported ``Expectation Misalignment'' (9.80\% vs 6.42\%) and issues with ``Tooling'' (5.88\% vs 1.07\%), which they attributed to managing the ``Volume'' of contributions (9.80\% vs 5.88\%) requiring reviews. Surprisingly, more Maintainers (2.94\%) than Contributors (1.07\%) identified ``Bias'' as an issue.

Maintainers frequently highlighted the difficulty of managing geographically distributed teams and coordinating asynchronous communication. For example, Participant M63 noted, ``\textit{Conversing with geographically distributed teams is challenging, especially when time zones and language barriers come into play.}'' These insights underscore the importance of fostering clear communication and setting expectations early in the review process.

\begin{tcolorbox}[colback=gray!10, colframe=gray!80, title=Key Finding]
Both groups cite ``Reviewer Responsiveness'' (29.07\%) as the top challenge, but Contributors emphasize ``Lack of Documentation'' (5.88\%) and ``Inconsistent Reviewer Experience'' (2.67\%), while Maintainers highlight ``Expectation Misalignment'' (9.80\%) and ``Tooling'' issues (5.88\%).
\end{tcolorbox}

\subsubsection*{\textbf{Improvements:}}
Contributors and Maintainers tend to agree more on potential improvements to the code review process than on its challenges. As shown in Table \ref{tab:improvements}, both groups emphasized the need for ``Better Tools and Automation'' (17.30\%), with more Contributors (18.72\%) advocating for this than Maintainers (14.71\%). This aligns with findings from follow-up interviews, where 16 participants highlighted the need for tools and automation to streamline the process. For example, Participant M4 suggested, ``More automated tools can help in addressing repetitive tasks such as checking code quality and running tests, freeing up maintainers to focus on higher-level review tasks.'' Participant M61 echoed this, stating, ``Automating repetitive tasks like code style checks would save a lot of time and reduce human errors.'' Participant M63 emphasized the potential of AI, noting, ``AI-powered tools could help identify common issues earlier in the review process and provide actionable suggestions.''

\begin{table}[htbp!]
    \caption{\small{Suggestions for Improving the Code Review Process}}
    \resizebox{0.45\textwidth}{!}{
        \begin{tabular}{|l|c|c|c|}
            \hline
            \textbf{Category} & \textbf{C (\%)} & \textbf{M (\%)} & \textbf{All (\%)} \\ \hline
            Better Tools and Automation    & 18.72                     & 14.71                     & 17.30             \\ \hline
            Improved Documentation         & 15.51                     & 11.76                     & 14.19             \\ \hline
            More Engagement                & 10.70                     & 6.86                      & 9.34              \\ \hline
            Miscellaneous                  & 6.42                      & 11.76                     & 8.30              \\ \hline
            Enhanced Communication         & 5.35                      & 12.75                     & 7.96              \\ \hline
            Improved Timeliness            & 4.81                      & 2.94                      & 4.15              \\ \hline
            Enhanced Collaboration         & 0.53                      & 0.98                      & 0.69              \\ \hline
        \end{tabular}
    }
    \label{tab:improvements}
    \small $C = Contributors;  M=Maintainers$
\end{table}
In response to challenges in the current asynchronous communication, some participants indicated the need for synchronous communication, e.g., M71 suggested ``\textit{It [should] be replaced as much as possible with synchronous collaboration such as mob programming.}''. However, this could be expensive and even might introduce more room for bias if the aim is anonymity during code review. 

These responses demonstrate a strong demand for technological advancements to improve efficiency and enhance the overall code review experience. Additionally, participants reported a need for ``Improved Documentation'' (14.19\%), further underscoring the importance of clear and accessible project guidelines.

One interesting observation from Table \ref{tab:improvements} is that 8.30\% of responses are too  diverse that they ended up in a miscellaneous category during the open coding.  More Maintainers (12.75\%) indicated the need for ``Enhanced Communication'' than Contributors (5.53\%) which was a statistically significant difference ($p-value < 0.05$). 

\begin{tcolorbox}[colback=gray!10, colframe=gray!80, title=Key Finding]
Both groups prioritize ``Better Tools and Automation'' (17.30\%) and ``Improved Documentation'' (14.19\%), but Maintainers significantly emphasize ``Enhanced Communication'' (12.75\% vs. 5.35\%).
\end{tcolorbox}

We asked participants to evaluate the code review process using 5-point Likert scale questions across four key areas: improving software quality, mitigating bias, promoting diversity and inclusivity, and adhering to communication standards to minimize conflict. Contributors and Maintainers largely aligned in their responses, though some variations were observed.


\textbf{Software Quality}
Participants rated this aspect positively, with 83.39\% agreeing or strongly agreeing that code reviews improve software quality. The mean rating was 4.21, with Contributors scoring slightly lower (mean = 4.19) than Maintainers (mean = 4.24). Both groups exhibited low variability, with standard deviations of 0.77 and 0.86, respectively. 

\textbf{Bias Mitigation}
Neutral responses dominated (43.25\%), with 30.10\% agreeing and 14.19\% rating negatively, indicating room for improvement. The mean was 3.28, with Contributors scoring slightly higher (mean = 3.29) than Maintainers (mean = 3.26). Variability was higher than for ``Software Quality,'' with standard deviations of 0.84 for Contributors and 0.99 for Maintainers.


\textbf{Diversity and Inclusivity}
Ratings leaned neutral or slightly positive, with 41.18\% selecting ``Neutral'' and 24.91\% agreeing. Negative responses (17.30\%) suggest challenges in this area. The mean was 3.25, with Contributors rating higher (mean = 3.27) than Maintainers (mean = 3.20), and variability was slightly greater among Maintainers.

\textbf{Communication Standards}
Responses were largely positive, with 61.93\% agreeing or strongly agreeing. The mean was 3.70, with Contributors scoring slightly higher (mean = 3.74) than Maintainers (mean = 3.62). Variability was low, though slightly higher for Maintainers.

\begin{tcolorbox}[colback=gray!10, colframe=gray!80, title=Key Finding]
Participants find the code review process effective for improving software quality and communication, but concerns about bias mitigation and inclusivity reveal areas for improvement, reflected in neutral and negative responses.
\end{tcolorbox}

\subsection{RQ2: Bias Experience}
We asked participants whether they had noticed bias in the code review process. The results indicate that 41.18\% of all respondents reported experiencing a biased code review at least once. When we analyze the responses by category, Maintainers (43.14\%) slightly surpass Contributors (40.11\%) by just over 3\%. One possible explanation for this difference is that Maintainers are typically developers who are more active in the open-source community compared to Contributors.

\begin{tcolorbox}[colback=gray!10, colframe=gray!80, title=Key Finding]
While less than 2\% of respondents indicated bias as a challenge, 41.18\% of reported to have witnessed/experienced bias in the code review process.
\end{tcolorbox}

In addition to Contributors vs Maintainers, we compared the responses across other demographic parameters. Previous studies show biases are normally experienced by minority groups \cite{sultana2023CodeReviews}, hence, we identified the dominant groups for each of the parameters we collected and compared difference in witnessing of bias between the dominant group and the rest which is presented in Table \ref{tab:bias}. We initially intended to analyze the data along the race/ethnicity dimension; however, a large number of participants (49.13\%) chose not to disclose their race/ethnicity or, rendering any meaningful analysis in this area impractical. As a result, we decided not to pursue this line of analysis. 

\begin{table}[htbp!]
    \centering
    \caption{Reported experience of bias in the code review process between most frequent group vs the rest.}
    \resizebox{0.45\textwidth}{!}{
        \begin{tabular}{|c|c|c|c|}
             \hline
             Category & Group & Noticed Bias & $p-value$\\
             \cline{1-4}
             \multirow{2}{*}{Age Group} & 25-34 & 43.66\% & \multirow{2}{*}{0.47}  \\
             \cline{2-3}
             & Others & 38.78\% & \\
             \cline{1-4}
    
             \multirow{2}{*}{Gender} & Male & 40.15\% & \multirow{2}{*}{0.29} \\
             \cline{2-3}
             & Others & 55.00\% & \\
             \cline{1-4}
    
             \multirow{2}{*}{Educational Level} & Bachelors & 39.39\% & \multirow{2}{*}{0.66}  \\
             \cline{2-3}
             & Others & 42.68\% & \\
             \cline{1-4}

             \multirow{2}{*}{Professional Experience} & 10< & 45.79\% & \multirow{2}{*}{0.27}  \\
             \cline{2-3}
             & Others & 38.46\% & \\
             \cline{1-4}
    
             \multirow{2}{*}{OSS Experience} & 4-6 & 41.18\% & \multirow{2}{*}{1.00} \\
             \cline{2-3}
             & Others & 41.18\% & \\
             \cline{1-4}

             \multirow{2}{*}{Current Responsibility} & Developer & 34.95\% & \multirow{2}{*}{0.0056} \\
             \cline{2-3}
             & Others & 52.43\% & \\
             \cline{1-4}         
        \end{tabular}
    }
    \label{tab:bias}
\end{table}
For the other parameters, we did not find any statistically significant difference other than self reported role in the projects. Participants whose primary responsibility was identified as ``Developer'' reported witnessing biased code reviews or interactions less frequently compared to those in other roles. 

\begin{tcolorbox}[colback=gray!10, colframe=gray!80, title=Key Finding]
Developers reported to have witnessed bias less compared to other roles.
\end{tcolorbox}

For participant that indicated to have witnessed bias, we asked them to elaborate on their experience. Their responses show that familiarity bias is the most frequently reported factor contributing to bias in the code review process, with 65.40\% of respondents. Familiarity bias encompasses responses that suggested Maintainers were unfairly favoring contributions from developers they knew beforehand, whether through previous collaborations or other social connections.\footnote{Full list of codes and description for each is found in the supplementary materials~\cite{supplemental}.}  


\begin{table}[h!]
    \centering
    \caption{Top five ``Biases'' witnessed by participants}
    \resizebox{0.45\textwidth}{!}{
        \begin{tabular}{|p{1cm}|p{.9cm}|p{3.75cm}|}
            \hline
            \footnotesize \textbf{Bias} & \footnotesize \textbf{Response} & \footnotesize \textbf{Description} \\ \hline
            \footnotesize Familiarity Bias & \small 65.40\% & \footnotesize Bias that occurred when the reviewer favored or was lenient towards familiar colleagues or their work. \\ \hline
            \footnotesize Approach Difference & \small 37.02\% & \footnotesize Bias resulting from differences in technical approaches or preferences between the reviewer and the author. \\ \hline
            \footnotesize Language Challenges & \small 16.61\% & \footnotesize Bias arising from difficulties in communication due to language differences. \\ \hline
            \footnotesize Misunder-standing & \small 8.65\% & \footnotesize Instances where bias was the result of miscommunication or misunderstanding.\\ \hline
            \footnotesize Other & \small 7.27\% & \footnotesize Bias-related situations that did not fit into specific categories but were still noted.  \\ \hline
        \end{tabular}
    }
    \label{tab:bias_types}
\end{table}

Approach differences were the second most cited experience of ``Bias'' (37.02\%) which refers to variations in design decision, coding style, techniques, or problem-solving methods preferred by developers, which are often based on personal experience, familiarity with certain frameworks, or individual interpretation of project requirements. However, this does not meet the definition of bias which is defined in the context of code review as unfair favoring or disfavoring of code contribution from certain individuals or groups based on characteristics unrelated to the quality of code, such as gender, race, or perceived experience level~\cite{Ford2019}. This indicates that there is a misunderstanding of bias which is inline with pervious studies \cite{Tsay2014}.


Language challenges, the third most reported bias overall (16.61 \%), is much more frequently identified by Maintainers (13.10\%) than Contributors (2.77\%), indicating a potential communication gap in OSS code reviews that Maintainers see as a source of unnecessary pushback. We classified all responses that indicated difficulties due to language barriers, such as miscommunication or misunderstandings caused by non-native language use, in this category. This is not a bias by itself, but could imply that contributions from these groups could be pushed back more than from developers who are native or fluent speakers of English.

\begin{tcolorbox}[colback=gray!10, colframe=gray!80, title=Key Finding]
Maintainers (24.10\%) reported more issues from communication barriers than Contributors (2.70\%), highlighting challenges for non-native English speakers in OSS code reviews.
\end{tcolorbox}

\subsection{RQ3: Documentations and Guidelines}


This section explores participants’ perceptions of project documentation related to the code review process. Using 6-point\footnote{The $6^{th}$ option is for not applicable. NA responses were removed for analysis.} Likert scale questions, we assessed the sufficiency, accessibility, clarity, and alignment of documentation with the objectives and challenges of the code review process. The results highlight strengths in documentation accessibility and alignment with expectations but reveal gaps in addressing bias and inclusivity. 

We began by asking respondents whether they consult project-specific documentation related to the code review process. Overall, 70\% of respondents indicated that they do. Among Contributors, 78.61\% reported consulting the documentation, compared to 54.90\% of Maintainers. We did not find any statistically significant difference between those that read documentation and not on witnessing bias. 

Participants generally rated documentation positively. Most agreed that guidelines for code review are sufficiently documented (51.28\% positive), with Maintainers rating slightly higher (mean = 3.84) than Contributors (mean = 3.67). Accessibility of documents was also well-rated, with 61.54\% responding positively. Expectations were moderately clear, with 48.68\% agreeing, though Contributors rated slightly lower (mean = 3.65) than Maintainers (mean = 3.81).

In terms of alignment between code review objectives and expectations, 50.57\% rated this positively, with Maintainers providing higher ratings (mean = 4.02) compared to Contributors (mean = 3.74). Neutral and negative responses (24.33\%) indicate some gaps in clearly communicating review objectives. Additionally, 9.00\% found the question inapplicable, underscoring potential discrepancies in understanding project goals.

Perceptions of bias-related documentation and processes were less favorable. Only 37.07\% agreed that documents effectively address bias, while 31.87\% were neutral, and 18.32\% rated negatively. Similarly, only 19.14\% agreed that bias mitigation processes are clearly defined, with notable negative responses (20.71\%). These findings highlight a need for improved guidance to address inclusivity and transparency in OSS projects.

Standards for communication were moderately well-rated, with 41.51\% agreeing they are clearly defined. Contributors and Maintainers showed minimal differences in their ratings across these areas, but the variability in responses, coupled with the percentage of participants marking some questions as inapplicable (ranging from 4.84\% to 13.15\%), suggests the need for clearer, more inclusive guidelines tailored to diverse project dynamics.

\begin{tcolorbox}[colback=gray!10, colframe=gray!80, title=Key Finding]
Participants find the documentation accessible and aligned with expectations, but gaps remain in addressing bias and defining mitigation processes, as reflected in neutral and negative responses.
\end{tcolorbox}

We analyzed responses based on whether participants reported consulting project-specific documentation. Respondents who read the documentation consistently rated aspects of the code review process more favorably. For instance, the mean rating for sufficiency of documentation was 3.82 among those who consulted the documentation, compared to 3.09 for those who did not. Similar trends were observed for accessibility (4.13 vs. 3.59), clarity of expectations (3.77 vs. 3.25), and alignment with expectations (3.88 vs. 3.42). Ratings for addressing bias (3.19 vs. 2.74) and defining communication standards (3.59 vs. 3.06) also showed notable differences. Lower variability in responses among those who read the documentation further suggests greater consensus. These results highlight the positive impact of consulting documentation on perceptions of clarity, alignment, and fairness in the code review process.


\section{Discussion}

Our findings indicate Contributors and Maintainers largely align on their perceptions of OSS code review processes, though subtle but significant differences exist. Maintainers prioritize project alignment (RQ1) and view existing documentation more positively (RQ3). Significant differences were observed in experiences of bias between developers and other roles (RQ2), and both groups emphasized the need for better tooling and documentation (RQ1). This section discusses key insights and broader implications.

\subsection{Importance of Project Directions}
Open-source projects are often viewed as community-driven~\cite{hippel}, with features and priorities evolving organically. However, core Maintainers play critical roles in guiding decisions, setting priorities, and balancing governance with open collaboration~\cite{Dabbish2012}. Their expertise ensures coherence, adherence to standards, and effective management of contributions.

Friction often arises from misaligned priorities between Maintainers and Contributors (see Table~\ref{tab:objectives}). While 15\% of Maintainers identified alignment with project goals as a primary objective, only 6.49\% of Contributors shared this view. As Maintainer M99 noted, “\textit{Contributions must align with the project’s direction. Unplanned features are unlikely to be approved}.”

Our follow-up interviews confirmed findings by Lerner and Tirole~\cite{lerner2002open}, showing developers often prioritize personal needs over project goals in initial contributions, leading to rejection and disengagement. For instance, Participant M65 abandoned a project after a pull request was rejected for misalignment with guidelines. This underscores the need for clearer onboarding and communication to help contributors align with project priorities.

Limited documentation and Maintainer-only communication channels deepen this disconnect, leaving contributors excluded from key decisions. Addressing these gaps through improved resources, as suggested by \cite{Jiang2013}, and conversational AI tools~\cite{dominic}, could better align contributors with project goals.

Future research should explore scalable methods to foster transparency in decision-making, assess the effectiveness of AI-based onboarding tools, and examine the dynamics of power imbalances in Maintainer-Contributor relationships to promote inclusivity and collaboration in open-source projects.

\subsection{Delay in Review Feedback}
A key challenge highlighted in our study is the delay in review feedback, affecting both Contributors and Maintainers. The limited number of active reviewers relative to the volume of contributions creates bottlenecks, especially in larger projects where reviewers juggle multiple responsibilities~\cite{Kononenko2016}. Availability and timely feedback are critical traits of effective OSS project Maintainers~\cite{Dias}.

Automated tools help alleviate delays by handling routine tasks like coding standard enforcement before human review~\cite{McIntosh2016}, reducing effort and response times~\cite{stolee}. As Participant M39 noted an automated processes could catch specific nuances, like added complexity, and would let reviewers focus on other aspects of the code. Automation also mitigates familiarity bias, ensuring fairer feedback.

Maintainers report using AI-based large language models (LLMs), such as ChatGPT\footnote{\url{https://chatgpt.com/}}, to support code review, though research shows perceptions remain mixed~\cite{devgpt, watanabe}. Future research may explore LLMs to improve review efficiency and reduce biases.

Expanding participation and accelerating the path to \textit{maintainership} will also help and addition to being crucial for OSS sustainability. Mentorship, structured onboarding, and inclusive governance improve contributor retention and maintainer progression~\cite{Steinmacher2015, Avelino2019}. Initiatives like ``community bonding'' and tiered contribution models further integrate newcomers into leadership roles, ensuring long-term project health.

\subsection{Familiarity Bias and Growing Communities}
Familiarity bias, the most frequently reported bias in our study, significantly hinders new Contributors (see Table~\ref{tab:bias_types}). This bias leads Maintainers and reviewers to favor contributions from known individuals, creating an uneven playing field. Contributor C76 observed, “\textit{There is bias toward known Contributors, whose work is accepted more readily, even when of equal or lower quality than that of new contributors}.” Maintainer M16 similarly noted, “\textit{PRs from familiar contributors are approved quickly without the same scrutiny}.” Such bias discourages new Contributors, reducing participation and engagement~\cite{Ford2019}.

Anonymized code reviews, which hide contributor identities, could help mitigate familiarity bias and ensure fairer evaluations, especially for newcomers. Both Contributors and Maintainers also highlighted the lack of active participants and reviewers as a persistent issue. With OSS contributions often being episodic and infrequent~\cite{barcomb}, discouraging new Contributors through bias further strains the limited contributor pool.

Addressing familiarity bias is crucial for fostering fairness, expanding the OSS community, and ensuring sustainable growth. Social integration, which transitions Contributors from ``outsiders'' to ``insiders''~\cite{justin}, is key to building inclusivity. Mentoring programs or blind review processes~\cite{Rigby2013} could create a more welcoming environment, reduce biases, and support long-term engagement.

Future research could focus on evaluating the effectiveness of anonymized review systems, understanding the dynamics of social integration in OSS, and designing scalable mentoring programs to support new Contributors while promoting fairness and inclusivity.

\subsection{Misunderstanding of Bias}
Misunderstanding approach differences as bias significantly impacts OSS code reviews. Variations in coding style or solutions are often perceived as bias, despite not meeting its formal definition of unfair prejudice. For example, C109 remarked, ``\textit{The reviewer is biased towards their coding style, resolved through discussing the pros and cons of their suggested approach}.'' Similarly, C6 reported adjusting their code to fit a gatekeeper’s preferences, even when they believed their solution was better.

While perceived bias is important to address, prior research indicates it is often overestimated in human judgment~\cite{pronin}. Misunderstandings in OSS frequently stem from differences in expectations rather than actual biases~\cite{Tsay2014}. Negative pushback, though infrequent, can lead to frustration and project abandonment~\cite{negative}. Improved communication and clearer alignment between Contributors and Maintainers are crucial. Well-documented \texttt{CONTRIBUTING.md} files that outline high-level goals and common barriers for newcomers can help reduce misunderstandings~\cite{contributing, doasido, shepherd}. Factors like unconscious bias~\cite{greenwald1995implicit} and language barriers exacerbate perceived inequities. Maintainers note that unclear communication fosters misunderstandings, often misinterpreted as bias.

Future research should explore scalable strategies for enhancing communication, including tools to clarify project expectations and reduce misaligned perceptions. Studies on how unconscious biases and language barriers affect OSS interactions could further address perceived inequities and improve inclusivity in code review practices.
\section{Limitations and Future Work}
\subsection{Limitations}

This study provides valuable insights into perceptions and expectations of developers in OSS code review processes but has several limitations. First, the focus on larger OSS communities may bias results, as smaller or niche projects with overlapping Contributor and Maintainer roles and less formalized processes were not fully captured. These findings may not generalize to all OSS communities or closed-source software.

What's more, participants were categorized by primary roles, but Maintainers may also act as Contributors in other projects, influencing their perceptions. While participants were instructed to focus on their Maintainer roles, this duality remains a limitation. Also, the sample size and diversity may not represent the full spectrum of OSS experiences. Responses were skewed toward experienced Contributors and Maintainers, with limited input from underrepresented groups or newer Contributors, potentially missing their unique challenges.

Finally, the data reflects OSS practices during a specific period. As tools, practices, and community dynamics evolve, these findings may lose relevance. Future studies should account for such changes to ensure continued applicability.

\subsection{Future Work}
Building on our findings, we plan to develop tools and automated systems to aid in the code review process. This include one for performing preliminary evaluation of pull requests providing feedback so that the wait time for Contributors is reduced and Maintainers will focus on making final decisions. Another tools is aimed at automatically detecting potentially biased code review. This tool will use machine learning algorithms to detect and flag biases in review comments or decision-making patterns. Additionally, we will create and test interventions such as bias-awareness training and guidelines for Contributors and Maintainers to evaluate their impact on reducing bias and fostering inclusivity. Expanding our research to include a more diverse range of projects and participants, particularly from underrepresented groups, will further deepen our understanding of bias in OSS communities.

\section{Conclusion}

This study examines shared goals and challenges in OSS code reviews, revealing disparities between Contributors and Maintainers. Key findings reveal the impact of familiarity bias and the misinterpretation of approach differences as bias. Participants called for better tools and automation for faster review time. There is also a need for clearer documentation in communicating general project direction/goals. The study contributes to OSS efficiency and inclusivity, recommending AI-driven tools, mentoring, and further research to foster equity and innovation in OSS code review processes.

\newpage
\bibliographystyle{ACM-Reference-Format}

\end{document}